\documentclass[12pt]{article}
\usepackage{graphicx}
\usepackage{color}

\def\hybrid{\topmargin 0pt      \oddsidemargin 0pt
        \headheight 0pt \headsep 0pt
       \voffset-1cm
        \textwidth 6.25in       
       \textheight 9.5in       
        \marginparwidth 0.0in
        \parskip 5pt plus 1pt   \jot = 1.5ex}
\catcode`\@=11
\def\marginnote#1{}

\newcount\hour
\newcount\minute
\newtoks\amorpm
\hour=\time\divide\hour by60
\minute=\time{\multiply\hour by60 \global\advance\minute by-\hour}
\edef\standardtime{{\ifnum\hour<12 \global\amorpm={am}%
        \else\global\amorpm={pm}\advance\hour by-12 \fi
        \ifnum\hour=0 \hour=12 \fi
        \number\hour:\ifnum\minute<10 0\fi\number\minute\the\amorpm}}
\edef\militarytime{\number\hour:\ifnum\minute<10 0\fi\number\minute}

\def\draftlabel#1{{\@bsphack\if@filesw {\let\thepage\relax
   \xdef\@gtempa{\write\@auxout{\string
      \newlabel{#1}{{\@currentlabel}{\thepage}}}}}\@gtempa
   \if@nobreak \ifvmode\nobreak\fi\fi\fi\@esphack}
        \gdef\@eqnlabel{#1}}
\def\@eqnlabel{}
\def\@vacuum{}
\def\draftmarginnote#1{\marginpar{\raggedright\scriptsize\tt#1}}

\def\draftlabel#1{{\@bsphack\if@filesw {\let\thepage\relax
   \xdef\@gtempa{\write\@auxout{\string
      \newlabel{#1}{{\@currentlabel}{\thepage}}}}}\@gtempa
   \if@nobreak \ifvmode\nobreak\fi\fi\fi\@esphack}
        \gdef\@eqnlabel{#1}}
\def\@eqnlabel{}
\def\@vacuum{}
\def\draftmarginnote#1{\marginpar{\raggedright\scriptsize\tt#1}}

\def\draft{\oddsidemargin -.5truein
        \def\@oddfoot{\sl preliminary draft \hfil
        \rm\thepage\hfil\sl\today\quad\militarytime}
        \let\@evenfoot\@oddfoot \overfullrule 3pt
        \let\label=\draftlabel
        \let\marginnote=\draftmarginnote
   \def\@eqnnum{(\theequation)\rlap{\kern\marginparsep\tt\@eqnlabel}%
\global\let\@eqnlabel\@vacuum}  }

\def\numberbysection{\@addtoreset{equation}{section}
        \def\theequation{\thesection.\arabic{equation}}}

\def\underline#1{\relax\ifmmode\@@underline#1\else
        $\@@underline{\hbox{#1}}$\relax\fi}

\def\titlepage{\@restonecolfalse\if@twocolumn\@restonecoltrue\onecolumn
     \else \newpage \fi \thispagestyle{empty}\c@page\z@
        \def\thefootnote{\fnsymbol{footnote}} }

\def\endtitlepage{\if@restonecol\twocolumn \else  \fi
        \def\thefootnote{\arabic{footnote}}
        \setcounter{footnote}{0}}  
\relax

\hybrid

\newfont{\Bbb}{msbm10 scaled 1\@ptsize00}
\newfont{\Bbbb}{msbm7 scaled 1\@ptsize00}
\newcommand{\CC}{\mbox{\Bbb C}}

\newcommand{\DDD}{\raise-1pt\hbox{$\mbox{\Bbbb D}$}}

\newcommand{\UUU}{\raise-1pt\hbox{$\mbox{\Bbbb U}$}}

\newcommand{\ZZ}{\mbox{\Bbb Z}}
\newcommand{\z}{\raise-1pt\hbox{$\mbox{\Bbbb Z}$}}

\newtheorem{predl}{Proposition}[section]

\def\beq{\begin{equation}}
\def\eeq{\end{equation}}
\def\p{\partial}

\def\square{\hfill
{\vrule height6pt width6pt depth1pt} \break \vspace{.01cm}}

\begin{document}

\begin{titlepage}

\title{QKZ-Ruijsenaars correspondence revisited}

\author{A.~Zabrodin\thanks{National Research University Higher School of Economics,
Russian Federation;
Skolkovo Institute of Science and Technology, 143026 Moscow, Russian Federation;
e-mail: zabrodin@itep.ru}
\and
 A. Zotov\thanks{Steklov Mathematical
Institute of Russian Academy of Sciences, Gubkina str. 8, 119991,
Moscow, Russia; ITEP, B.Cheremushkinskaya 25, Moscow 117218, Russia;
Moscow Institute of Physics and Technology, Inststitutskii per.  9,
Dolgoprudny, Moscow region, 141700, Russia; e-mail: zotov@mi.ras.ru}
}

\date{April 2017}
\maketitle

\vspace{-7cm} \centerline{ \hfill ITEP-TH-04/17}\vspace{7cm}

\begin{abstract}

We discuss the Matsuo-Cherednik type correspondence between the
quantum Knizhnik-Zamolodchikov equations associated with $GL(N)$ and the
$n$-particle quantum Ruijsenaars model, with $n$ being not
necessarily equal to $N$. The quasiclassical limit of this
construction yields the quantum-classical correspondence between the
quantum spin chains and the classical Ruijsenaars models.

\end{abstract}

\end{titlepage}

\vspace{5mm}

\section{Introduction}

The quantum Knizhnik-Zamolodchikov (qKZ) equations \cite{FR}
is a system of holonomic difference equations
 \beq\label{qkz1}
e^{\eta \hbar \p_{x_i}} \Bigl | \Phi \Bigr >={\bf K}_i^{(\hbar )}
\Bigl | \Phi \Bigr >, \qquad i=1, \ldots , n
 \eeq
for the vector
$\Bigl | \Phi \Bigr >=\Bigl | \Phi \Bigr >(x_1, \ldots , x_n)$ belonging to the
tensor product ${\cal V}=V\otimes V\otimes \ldots \otimes V =V^{\otimes n}$ of the
vector spaces $V=\CC^N$. The operator ${\bf K}_i^{(\hbar )}$ in the right hand side
is constructed as a chain product of quantum $R$-matrices:
 \beq\label{qkz1a}
{\bf K}_i^{(\hbar )}=
{\bf R}_{i\, i\! -\! 1}(x_i \! -\! x_{i-1}\! +\! \eta \hbar )\ldots
{\bf R}_{i 1}(x_i \! -\! x_{1}\! +\! \eta \hbar ) {\bf g}^{(i)}
{\bf R}_{i n}(x_i \! -\! x_{n})\ldots
{\bf R}_{i\, i\! +\! 1}(x_i \! -\! x_{i+1} ).
 \eeq
Here
${\bf R}_{ij}(x)$ is the $R$-matrix acting in the $i$-th and $j$-th
tensor factors (it has to satisfy the unitarity condition
${\bf R}_{ij}(x){\bf R}_{ji}(-x)=\mbox{id}$), ${\bf g}=\mbox{diag}(g_1, \ldots , g_N)$
is a diagonal $N\! \times \! N$ matrix and
${\bf g}^{(i)}$ is the operator in ${\cal V}$ acting as ${\bf g}$ on the $i$-th factor
(and identically on all other factors). For example, the rational
$R$-matrix is of the form
 \beq\label{r1}
{\bf R}_{ij}(x)=\frac{x{\bf I} +\eta {\bf P}_{ij}}{x+\eta},
 \eeq
where $x$ is the spectral parameter, ${\bf I}$ is the identity operator and ${\bf P}_{ij}$
is the permutation of the $i$-th and $j$-th
tensor factors. Compatibility of the qKZ equations follows from the
Yang-Baxter equation for the $R$-matrix and from the commutativity
$[{\bf g}\otimes {\bf g}, \, {\bf R}(x)]=0$.

The remarkable correspondence of the qKZ equations with the
Macdonald type difference operator
 \beq\label{mac1}
\hat {\cal H}=\sum_{i=1}^n  \left (\prod_{j\neq i}^n
\frac{x_i-x_j +\eta}{x_i -x_j}\right ) e^{\eta \hbar \p_{x_i}}
 \eeq
was discussed in \cite{Cherednik1,Kato,Mimachi,Stokman} in the case
$N=n$. In this case solutions to (\ref{qkz1}) can be found in the form
 $$
\Bigl | \Phi \Bigr >=\sum_{\sigma \in S_n}\Phi_{\sigma}\Bigl | e_{\sigma} \Bigr >, \qquad
\Bigl | e_{\sigma} \Bigr >=e_{\sigma (1)}\otimes e_{\sigma (2)}\otimes \ldots \otimes
e_{\sigma (n)},
 $$
where $e_{a}$ are standard basis vectors in $\displaystyle{V=\CC^N=\bigoplus _{a=1}^{N}\CC e_a}$
and $S_n$ is the symmetric group.
If such $\Bigl | \Phi \Bigr >$ is a solution to the qKZ equations,
then the function
$\displaystyle{
\Psi =\sum_{\sigma \in S_n}\Phi_{\sigma}}
$
solves the spectral problem for the operator $\hat {\cal H}$:
 \beq
\label{mac3}
\sum_{i=1}^n  \prod_{j\neq i}^n
\frac{x_i\! -\! x_j \! +\! \eta}{x_i -x_j}\, \Psi (x_1, \ldots , x_i\! +\! \eta \hbar ,
\ldots , x_n) =E\Psi (x_1, \ldots , x_n) , \qquad E=\sum_{a=1}^N g_a.
 \eeq
This is referred to as the Matsuo-Cherednik type correspondence \cite{Matsuo,Cherednik}.
A similar correspondence holds true for qKZ equations with trigonometric $R$-matrices.

The operator $\hat {\cal H}$ is essentially the Hamiltonian of the
quantum Ruijsenaars model \cite{Ruijsenaars} which is a relativistic
version of the Calogero model (more precisely, the operator $\hat {\cal H}$ and
the Ruijsenaars Hamiltonian are connected by
a similarity transformation, see \cite{H}). The parameter $\eta$ plays the role of the
inverse velocity of light.

We will give a simple proof of the correspondence valid also in the case when
$n$ is not necessarily equal to $N=\mbox{dim}\, V$. In this form it looks like
a quantum deformation of the quantum-classical correspondence
\cite{AKLTZ13,BLZZ16,GZZ14,Z2,BG} between the
quantum XXX or XXZ spin chains and the classical Ruijsenaars models.

The spectral problem for the spin chain appears as a ``quasiclassical'' limit
of qKZ as $\hbar \to 0$. Indeed, as $\hbar \to 0$ the
qKZ solutions have the asymptotic form \cite{TV1,TV2}
 $$
\Bigl |\Phi \Bigr >=\left ( \Bigl |\phi_0\Bigr > + \hbar \Bigl |\phi_1\Bigr >
+\ldots \right )
e^{S/\hbar},
 $$
where $S$ is some scalar function.
Upon substitution to the qKZ equations (\ref{qkz1}), this leads, in the leading order,
to the
joint eigenvalue problems
 \beq\label{quasi1}
{\bf K}_i^{(0)} \Bigl |\phi_0\Bigr >=e^{\eta p_i} \Bigl |\phi_0\Bigr >, \qquad
p_i =\frac{\p S}{\p x_i},\qquad i=1, \ldots , n,
 \eeq
for the commuting operators ${\bf K}_i^{(0)}$. They are Hamiltonians of the inhomogeneous
quantum spin chain. They can be diagonalized using the algebraic Bethe ansatz.
In the quan\-tum-clas\-si\-cal correspondence, the $p_i$'s are identified with
momenta of the Ruijsenaars particles.
The parameter $\hbar$ becomes
the true Planck constant after quantization of the corresponding Ruijsenaars model.

The paper is organized as follows. In section 2, we review the inhomogeneous XXX spin chain
and the associated qKZ equations. In section 3
the qKZ-Ruijsenaars correspondence is established by a simple direct calculation.
In section 4 we extend the result to higher Ruijsenaars Hamiltonians.
Section 5 is devoted to the qKZ equations with trigonometric $R$-matrices.
Finally, in section 6
we discuss the interpretation of the results
as a ``quantum'' deformation of the quantum-classical correspondence.
The link to spin chains and their solutions by means of the
algebraic Bethe ansatz appears naturally because of the usage of
$R$-matrices in fundamental representation. It is alternative to the
 approach based on the affine Hecke algebras
\cite{Cherednik,Cherednik1,Cherednik2}, where similar results were
originally presented. Explicit relationship between these two
approaches deserves further consideration.

\section{The XXX spin chain and qKZ equations}

Let $e_{ab}$ be the standard
basis in the space of $N\! \times \! N$ matrices:
the matrix $e_{ab}$ has only
one non-zero element (equal to 1) at the place $ab$:
$(e_{ab})_{a'b'}=\delta_{aa'}\delta_{bb'}$.
Note that ${\bf I}=\sum_a e_{aa}$ is the unity operator and
${\bf P}=\sum_{ab}e_{ab} \otimes e_{ba}$ is the permutation operator
in the space $\CC^N \otimes \CC^N$.
We embed
$e_{ab}$ into $\mbox{End}\, (V^{\otimes n})$
in the usual way:
$e^{(i)}_{ab}:= {\bf I}^{\otimes (i-1)} \otimes
e_{ab} \otimes  {\bf I}^{\otimes (n-i)}$.
It is clear that $e^{(i)}_{ab}$,  $e^{(j)}_{a'b'}$ commute for  any $i \ne j$
because they act non-trivially in different spaces.
Similarly, for any matrix ${\bf g}\in \mbox{End}(\CC^N)$ we
define ${\bf g}^{(i)}$ acting in the tensor product
$V^{\otimes n}$:
${\bf g}^{(i)}= {\bf I}^{\otimes (i-1)} \otimes
{\bf g}\otimes  {\bf I}^{\otimes (n-i)}\in \mbox{End}({\cal V})$.
In this notation, the
permutation operator of the $i$-th and $j$-th tensor factors
in ${\cal V}=\CC^N \otimes \ldots \otimes \CC^N$ is
$\displaystyle{{\bf P}_{ij}=\sum_{a,b}e^{(i)}_{ab}e^{(j)}_{ba}}$.

Let ${\bf g}\in GL(N)$ be a diagonal matrix
$\displaystyle{{\bf g}=\mbox{diag}\, (g_1, g_2, \ldots , g_N)=\sum_{a=1}^Ng_a e_{aa}}$.
We call it the twist matrix with twist parameters $g_a$.
It is used for the construction of an integrable spin chain
with twisted boundary conditions.
Together with the $R$-matrix (\ref{r1}), we introduce another rational
$R$-matrix which differs from the ${\bf R}(x)$ by a scalar factor:
 \beq\label{r2}
\widetilde {\bf R}(x)=
\frac{x+\eta}{x}\, {\bf R}(x) ={\bf I}+\frac{\eta}{x}\, {\bf P}.
 \eeq
It obeys the same Yang-Baxter equation and commutes with ${\bf g}\otimes {\bf g}$.
The transfer matrix of the inhomogeneous spin chain
(or, equivalently, of the associated statistical vertex model on the 2D lattice) is defined
in the standard way as a trace of the chain product of $R$-matrices
in the auxiliary space $V=\CC ^N$ with index $0$:
 \beq\label{r3}
{\bf T}(x)=\mbox{tr}_0\Bigl (\widetilde {\bf R}_{0n}(x-x_n)\ldots
\widetilde {\bf R}_{02}(x-x_2) \widetilde {\bf R}_{01}(x-x_1)({\bf
g}\otimes {\bf I})\Bigr ),
 \eeq
where $x_1, x_2, \ldots , x_n$ are inhomogeneity parameters. (We assume that they are in
general position meaning that $x_i\neq x_j$ and $x_i\neq x_j \pm \eta$ for all $i\neq j$.)
As is known, the Yang-Baxter equation for the $R$-matrix implies that the transfer matrices
with fixed inhomogeneity and twist parameters
commute: $[{\bf T}(x), \, {\bf T}(x')]=0$.

The dynamical variables of the model (we call them ``spins'' in analogy with the rank 1 case)
are vectors in the vector representation of $GL(N)$ realized in the spaces
$V=\CC ^N$ attached to each site. Non-local commuting Hamiltonians ${\bf H}_j$ are defined as
residues of ${\bf T}(x)$ at $x=x_j$:
 \beq\label{r4}
{\bf T}(x)=\mbox{tr} \,{\bf g}\cdot {\bf I}+\sum_{j=1}^n\frac{\eta {\bf H}_j}{x-x_j}.
 \eeq
From (\ref{r3}) it follows that their explicit form is
 \beq\label{r6}
{\bf H}_i=
\widetilde {\bf R}_{i\, i\! - \! 1}(x_i \! -\! x_{i-1})\ldots
\widetilde {\bf R}_{i 1}(x_i \! -\! x_{1}) {\bf g}^{(i)}
\widetilde {\bf R}_{i n}(x_i \! -\! x_{n})\ldots
\widetilde {\bf R}_{i\, i\! +\! 1}(x_i \! -\! x_{i+1} ).
 \eeq
We obviously have
 \beq\label{r5}
{\bf H}_i={\bf K}_i^{(0)}\prod_{j\neq i}^n \frac{x_i-x_j+\eta}{x_i-x_j},
 \eeq
where ${\bf K}_i^{(0)}$ is the operator (\ref{qkz1a}) at $\hbar =0$.

Let us introduce the operators
 \beq\label{Ma}
{\bf M}_a=\sum_{l=1}^{n}e_{aa}^{(l)}.
 \eeq
They commute among themselves and with the
Hamiltonians: $[{\bf M}_a, {\bf M}_b]=[{\bf H}_i, {\bf M}_a]=0$.
We call them weight operators.
Clearly,
$\sum_a {\bf M}_a =n {\bf I}$.
Comparing the expansion of (\ref{r3}) as $x\to \infty$,
 $$
\begin{array}{lll}
{\bf T}(x)&=&\displaystyle{\mbox{tr}_0 \left [ \Bigl (
{\bf I}+\frac{\eta {\bf P}_{0n}}{x-x_n}\Bigr )\ldots
\Bigl ({\bf I}+\frac{\eta {\bf P}_{01}}{x-x_1}\Bigr ){\bf g}^{(0)}\right ]}
\\ &&\\
&=&\displaystyle{\mbox{tr}\, {\bf g} \cdot {\bf I}+\frac{\eta}{x}\sum_{i=1}^n
\mbox{tr}_0 \Bigl ( {\bf P}_{0i}{\bf g}^{(0)}\Bigr )+\ldots}
\\ &&\\
&=&\displaystyle{\mbox{tr}\, {\bf g} \cdot {\bf I}+\frac{\eta}{x}\sum_{i=1}^n
{\bf g}^{(i)}+\ldots}\, ,
\end{array}
 $$
with that of (\ref{r4}),
we conclude that
 \beq\label{r7}
\sum_{i=1}^{n}{\bf H}_i= \sum_{i=1}^n {\bf g}^{(i)}=
\sum_{a=1}^{N}g_a {\bf M}_a,
 \eeq
so the system has $n+N-2$ independent integrals of motion.
The joint spectral problem is
 $$
\left \{\begin{array}{l}
{\bf H}_i \Bigl |\phi \Bigr > =H_i \Bigl |\phi \Bigr >
\\
{\bf M}_a \Bigl |\phi \Bigr > =M_a \Bigl |\phi \Bigr >
\end{array}
\right.
 $$
The common eigenstates of the Hamiltonians can be classified
according to eigenvalues of the operators ${\bf M}_a$.

Let
 \beq\label{weight1}\displaystyle{
{\cal V}= V^{\otimes n} \,\, =
\bigoplus_{M_1, \ldots , M_N} \!\! \!\! {\cal V}(\{M_a \})}
 \eeq
be the weight decomposition of the Hilbert space of the spin chain,
${\cal V}$, into the direct sum of eigenspaces for the operators
${\bf M}_a$ with the eigenvalues $M_a \in \ZZ_{\geq 0}$, $a=1, \ldots , N$
(recall that $M_1 +\ldots +M_N=n$).
The common eigenstates of the operators
${\bf M}_a$ and ${\bf H}_i$ belong to the spaces ${\cal V}(\{M_a \})$. The dimension
of ${\cal V}(\{M_a \})$ is
 $$
\mbox{dim} {\cal V}(\{M_a \})=\frac{n!}{M_1! \ldots M_N!}.
 $$
The basis vectors in ${\cal V}(\{M_a \})$ are
$
\Bigl |J\Bigr > =e_{j_1}\otimes e_{j_2}\otimes \ldots \otimes e_{j_n},
$
where the number of indices $j_k$ such that $j_k =a$ is equal to $M_a$ for all
$a=1, \ldots , N$.
We also introduce dual vectors $\Bigl <J\Bigr |$ such that
$\Bigl <J\Bigr |J'\Bigr >=\delta_{J, J'}$. Matrix elements of an operator
${\bf O}$ are $\Bigl <J\Bigr |{\bf O}\Bigl |J'\Bigr >$.

Associated with the inhomogeneous XXX spin chain is the system
of qKZ equations $e^{\eta \hbar \p_{x_i}} \Bigl | \Phi \Bigr >={\bf K}_i^{(\hbar )}
\Bigl | \Phi \Bigr >$ (\ref{qkz1}) with the operators ${\bf K}_{i}^{(\hbar )}$
given by (\ref{qkz1a}). The compatibility condition
 $$
\Bigl (e^{\eta \hbar \p_{x_i}}{\bf K}_{j}^{(\hbar )}\Bigr )
{\bf K}_{i}^{(\hbar )}=
\Bigl (e^{\eta \hbar \p_{x_j}}{\bf K}_{i}^{(\hbar )}\Bigr )
{\bf K}_{j}^{(\hbar )}
 $$
follows from the Yang-Baxter equation for the $R$-matrix.
The operators ${\bf K}_{i}^{(\hbar )}$
respect the weight decomposition (\ref{weight1}), hence the solutions
to the qKZ system belong to the weight subspaces ${\cal V}(\{M_a \})$.

\section{The qKZ-Ruijsenaars correspondence in the rational case}

Let $\displaystyle{\Bigl |\Phi \Bigr >=\sum_J \Phi_J \Bigl |J\Bigr >}$
be any solution of the qKZ equations belonging to the weight subspace
${\cal V}(\{M_a \})$.
We claim that
the function
 \beq\label{qkz2}
\Psi =\sum_J \Phi_J
 \eeq
is an eigenfunction of the Macdonald operator $\hat {\cal H}$ with the eigenvalue
$\displaystyle{E=\sum_{a=1}^N M_a g_a}$:
 \beq\label{qkz3}
\sum_{i=1}^n  \prod_{j\neq i}^n
\frac{x_i\! -\! x_j \! +\! \eta}{x_i -x_j}\, \Psi (x_1, \ldots , x_i\! +\! \eta \hbar ,
\ldots , x_n) =E\Psi (x_1, \ldots , x_n).
 \eeq

For the proof we consider the covector $\Bigl < \Omega \Bigr |$
equal to the sum of all basis (dual) vectors:
 $$
\Bigl < \Omega \Bigr |=\sum_J \Bigl < J \Bigr |,
 $$
then $\Psi =\Bigl <\Omega \Bigr | \Phi \Bigr >$. It is important to note that
$\Bigl <\Omega \Bigr |{\bf P}_{ij}=\Bigl <\Omega \Bigr |$ and, therefore,
$\Bigl <\Omega \Bigr |{\bf R}_{ij}(x)=\Bigl <\Omega \Bigr |$.
It then follows that $\Bigl <\Omega \Bigr |{\bf K}_{i}^{(\hbar )}=
\Bigl <\Omega \Bigr |{\bf K}_{i}^{(0 )}$, so the projection of the $i$-th qKZ equation
onto the covector $\Bigl <\Omega \Bigr |$ reads
 $$
e^{\eta \hbar \p_{x_i}}\Bigl <\Omega \Bigr |\Phi \Bigr >=
e^{\eta \hbar \p_{x_i}}\Psi =
\Bigl <\Omega \Bigr |   {\bf K}_{i}^{(\hbar )}\Bigl | \Phi \Bigr >=
\Bigl <\Omega \Bigr |   {\bf K}_{i}^{(0 )}\Bigl | \Phi \Bigr >.
 $$
Therefore, multiplying by $\displaystyle{\prod_{j\neq i}^n \frac{x_i-x_j +\eta}{x_i-x_j}}$
and summing over $i$, we get:
 $$
\sum_{i=1}^n \left ( \prod_{j\neq i}^n \frac{x_i-x_j +\eta}{x_i-x_j}\right )
e^{\eta \hbar \p_{x_i}}\Psi \,\, =\,\,
\sum_{i=1}^n  \prod_{j\neq i}^n \frac{x_i-x_j +\eta}{x_i-x_j}\,
\Bigl <\Omega \Bigr |   {\bf K}_{i}^{(0 )}\Bigl | \Phi \Bigr >
 $$
 $$
=\, \sum_{i=1}^n \Bigl <\Omega \Bigr |   {\bf H}_{i}\Bigl | \Phi \Bigr > \,\, =
\,\, \sum_{i=1}^n \Bigl <\Omega \Bigr |   {\bf g}^{(i)}\Bigl | \Phi \Bigr >\,\, =\,\,
\sum_{a=1}^N g_a \Bigl <\Omega \Bigr |   {\bf M}_a\Bigl | \Phi \Bigr >\,\, =\,\,
\left (\sum_{a=1}^N g_a M_a\right )\Psi ,
 $$
where we have used (\ref{r5}) and (\ref{r7}).

In the next section we show that
$\Psi$ is the common eigenfunction for all
higher Ruijsenaars Hamiltonians $\hat {\cal H}_d$ with the eigenvalues
$e_d(\underbrace{g_1, \ldots , g_1}_{M_1},\, \ldots \,
\underbrace{g_N, \ldots , g_N}_{M_N})$, where
$e_d$ is the elementary symmetric polynomial of $n$ variables.

\section{Higher Hamiltonians}

Here we show that $\Psi$ (\ref{qkz2}) is an eigenfunction of the
higher rational Mac\-do\-nald-\-Ruij\-sena\-ars Hamiltonians $\hat {\cal H}_d$ defined
by
 \beq\label{a1}
  \displaystyle{
  \hat {\cal H}_d=\sum\limits_{I\subset\{1, \ldots , n\}, |I|=d}\Bigl(
  \prod\limits_{s\in I,r\not\in I}\frac{x_s-x_r+\eta}{x_s-x_r}\Bigr)
  \prod\limits_{i\in I}e^{\eta\hbar\p_{x_i}}
  }
 \eeq
 or
  \beq\label{a2}
  \displaystyle{
\hat {\cal H}_d=\sum_{1 \le i_{1} < \ldots < i_{d} \le n}\
 \prod\limits_{k=1}^d\ \prod\limits_{r\neq i_k}^n\frac{x_{i_k}-x_r+\eta}{x_{i_k}-x_r}
\prod_{1 \le \alpha < \beta \le d} \left(
1-\frac{\eta^{2}}{(x_{i_{\alpha}}- x_{i_{\beta}})^{2} } \right)^{-1}
\prod\limits_{k=1}^d e^{\eta\hbar\frac{\p\ }{\p x_{i_k}}}\,.
  }
 \eeq
For example, for $d=2$
 \beq\label{a21}
 \begin{array}{c}
  \displaystyle{
  \hat {\cal H}_2=\sum\limits_{i<j}^n\Bigl(
  \prod\limits_{k\neq i,j}^n\frac{x_i-x_k+\eta}{x_i-x_k}\prod\limits_{l\neq
  i,j}^n\frac{x_j-x_l+\eta}{x_j-x_l}\Bigr)\,
  e^{\eta\hbar\p_{x_i}}e^{\eta\hbar\p_{x_j}}=
  }
  \\ \ \\
\displaystyle{
 =\sum\limits_{i<j}^n\Bigl(
  \prod\limits_{k\neq i}^n\frac{x_i-x_k+\eta}{x_i-x_k}\prod\limits_{l\neq j}^n\frac{x_j-x_l+\eta}{x_j-x_l}\Bigr)
  \left(
1-\frac{\eta^{2}}{(x_i- x_j)^{2} } \right)^{-1}
e^{\eta\hbar\p_{x_i}}e^{\eta\hbar\p_{x_j}}\,.
  }
  \end{array}
 \eeq

 \begin{predl}
 The operators ${\bf K}_j^{(\hbar)}$ from (\ref{qkz1}) and the wave function
 (\ref{qkz2}) $\Psi=\Bigl<\Omega | \Phi \Bigr>$  satisfy
 \beq\label{a3}
  \displaystyle{
\prod\limits_{s=1}^d e^{\eta\hbar\frac{\p\ }{\p x_{i_s}}}\Psi=\Bigl
<\Omega\Bigr |
 {\bf K}_{i_1}^{(0 )}\ldots{\bf K}_{i_d}^{(0  )} \Bigr | \Phi \Bigr
 >\quad \hbox{for}\quad i_k\neq i_m\,.
  }
 \eeq
 \end{predl}
 For the proof we introduce the notation ${\bf R}_{ab}={\bf R}_{ab}(x_a-x_b)$ and
 ${\bf R}^+_{ab}={\bf R}_{ab}(x_a-x_b+\eta\hbar)$, so that
$$
{\bf K}_j^{(\hbar)}={\bf R}_{jj-1}^+\ldots{\bf R}_{j1}^+\,{\bf
g}^{(j)}\,{\bf R}_{jn}\ldots {\bf R}_{jj+1}\,.
$$
Consider first the case $d=2$. For $i<j$ we have
 \beq\label{a4}
   \begin{array}{l}
   \displaystyle{
 e^{\eta\hbar\p_{x_i}}e^{\eta\hbar\p_{x_j}}\Psi=\Bigl <\Omega \Bigr |{\bf K}_j^{(\hbar
 )}(x_i+\eta\hbar){\bf K}_i^{(\hbar )}\Bigl |\Phi \Bigr >
 }
 \\ \ \\
  \displaystyle{
 =\Bigl <\Omega\Bigr |\, {\bf R}_{j\, j\! -\! 1}^+\ldots {\bf
R}_{j\, i\! +\! 1}^+ {\bf R}_{j\, i} {\bf R}_{j\, i\! -\!
1}^+\ldots{\bf R}_{j 1}^+\,
 \underbrace{{\bf
g}^{(j)}\, {\bf R}_{j n}\ldots {\bf R}_{j\, j\! +\! 1}}
  }
   \\ \ \\
  \displaystyle{ \times
\underbrace{{\bf R}_{ii-1}^+\ldots{\bf R}_{i1}^+\,{\bf
g}^{(i)}}\,{\bf R}_{in}\ldots {\bf R}_{ii+1}\,\Bigr | \Phi \Bigr
>\,.
  }
  \end{array}
 \eeq
 The ``underbraced'' expressions in (\ref{a4}) consist of commuting
 products of
 $R$-matrices thanks to $i<j$. Therefore, these expressions can be
 permuted. Using then the property $\Bigl <\Omega\Bigr |\,{\bf
 R}_{kl}(x)=\Bigl <\Omega\Bigr |$ we conclude that all shifts by $\eta\hbar$ in the $R$-matrix
 arguments can be removed. Finally, we can permute the products of $R$-matrices coming back to the
 initial order
 (but with non-shifted arguments).

 For $d=3$ and $i<j<k$ we have
 \beq\label{a41}
   \begin{array}{l}
   \displaystyle{
 e^{\eta\hbar\p_{x_i}}e^{\eta\hbar\p_{x_j}}e^{\eta\hbar\p_{x_k}}\Psi=\Bigl <\Omega \Bigr |{\bf K}_k^{(\hbar
 )}(x_i+\eta\hbar,x_j+\eta\hbar)
 {\bf K}_j^{(\hbar )}(x_i+\eta\hbar){\bf K}_i^{(\hbar )}\Bigl |\Phi \Bigr >
 }
 \\ \ \\
  \displaystyle{=
 \Bigl <\Omega\Bigr |\, {\bf R}_{k\, k\! -\! 1}^+\ldots{\bf
R}_{k\, j\! +\! 1}^+ {\bf R}_{k\, j} {\bf R}_{k\, j\! -\! 1}^+\ldots
{\bf R}_{k\, i\! +\! 1}^+ {\bf R}_{k\, i} {\bf R}_{k\, i\! -\!
1}^+\ldots{\bf R}_{k 1}^+\,
 \overbrace{{\bf
g}^{(k)}\, {\bf R}_{k n}\ldots {\bf R}_{k\, k\! +\! 1}}
  }
   \\ \ \\
  \displaystyle{ \times
 \overbrace{ {\bf R}_{j\, j\! -\! 1}^+\ldots {\bf
R}_{j\, i\! +\! 1}^+ {\bf R}_{j\, i} {\bf R}_{j\, i\! -\!
1}^+\ldots{\bf R}_{j 1}^+}\,
 \underbrace{{\bf
g}^{(j)}\, {\bf R}_{j n}\ldots {\bf R}_{j\, j\! +\! 1}}
  }
   \\ \ \\
  \displaystyle{ \times
\underbrace{{\bf R}_{ii-1}^+\ldots{\bf R}_{i1}^+\,{\bf
g}^{(i)}}\,{\bf R}_{in}\ldots {\bf R}_{ii+1}\,\Bigr | \Phi \Bigr
>\,.
  }
  \end{array}
 \eeq
Again, since $i<j<k$ we see that ``underbraced'' expressions commute as
well as the ``overbraced'' expressions. Permute first the
``underbraced'' expressions. Then we get
 \beq\label{a42}
   \begin{array}{l}
   \displaystyle{
 e^{\eta\hbar\p_{x_i}}e^{\eta\hbar\p_{x_j}}e^{\eta\hbar\p_{x_k}}\Psi=\Bigl < \Omega \Bigr |{\bf K}_k^{(\hbar
 )}(x_i+\eta\hbar,x_j+\eta\hbar)
 {\bf K}_j^{(\hbar )}(x_i+\eta\hbar){\bf K}_i^{(\hbar )}\Bigl |\Phi \Bigr >
 }
 \\ \ \\
  \displaystyle{=
 \Bigl <\Omega\Bigr |\, {\bf R}_{k\, k\! -\! 1}^+\ldots{\bf
R}_{k\, j\! +\! 1}^+ {\bf R}_{k\, j} {\bf R}_{k\, j\! -\! 1}^+\ldots
{\bf R}_{k\, i\! +\! 1}^+ {\bf R}_{k\, i} {\bf R}_{k\, i\! -\!
1}^+\ldots{\bf R}_{k 1}^+\,
 \overbrace{{\bf
g}^{(k)}\, {\bf R}_{k n}\ldots {\bf R}_{k\, k\! +\! 1}}
  }
   \\ \ \\
  \displaystyle{ \times
 \overbrace{ {\bf R}_{j\, j\! -\! 1}^+\ldots {\bf
R}_{j\, i\! +\! 1}^+ {\bf R}_{j\, i} {\bf R}_{j\, i\! -\!
1}^+\ldots{\bf R}_{j 1}^+}\, \underbrace{{\bf R}_{ii-1}^+\ldots{\bf
R}_{i1}^+\,{\bf g}^{(i)}}
  }
   \\ \ \\
  \displaystyle{\times
\underbrace{{\bf g}^{(j)}\, {\bf R}_{j n}\ldots {\bf R}_{j\, j\! +\!
1}}\,{\bf R}_{in}\ldots {\bf R}_{ii+1}\,\Bigr | \Phi \Bigr
>\,.
  }
  \end{array}
 \eeq
Now the ``overbraced'' expression from the second line of (\ref{a42})
commutes with the whole third line. By permuting them we get the product
where all $R$-matrices with shifted arguments ($\bf R^+$) are to the
left from twist matrices and act on $\Bigl <\Omega\Bigr |$ (to the
left). Then we can apply the previously used reasoning and remove
all the shifts of arguments.

It is easy to see that the same proof holds true for arbitrary $d$.
The choice of ordering $i_1<...<i_d$ is convenient but the final
answer is independent of it since $ [{\bf K}_i^{(0 )}, {\bf K}_j^{(0
)}]=0$ (because ${\bf K}_i^{(0 )}\propto \bf H_i$). $\square$

 Multiplying both parts of (\ref{a3}) by the products transforming ${\bf R}(x)$ to
 $\widetilde {\bf R}(x)$ (\ref{r2}), (\ref{r5}), we get
 \beq\label{a5}
  \displaystyle{
 \prod\limits_{k=1}^d\ \prod\limits_{r\neq i_k}^n\frac{x_{i_k}-x_r+\eta}{x_{i_k}-x_r}
\prod\limits_{k=1}^d e^{\eta\hbar\frac{\p\ }{\p x_{i_k}}}\Psi=\Bigl
<\Omega\Bigr |
 {\bf H}_{i_1}\ldots {\bf H}_{i_d} \Bigr | \Phi \Bigr >
  }
 \eeq
 A comparison between (\ref{a2}) and (\ref{a5}) shows that
 \beq\label{a6}
  \displaystyle{
\hat {\cal H}_d\Psi=\sum_{1 \le i_{1} < \ldots < i_{d} \le n}
 <\Omega\Bigr |{\bf H}_{i_{1}}\dots {\bf H}_{i_{d}} \Bigr | \Phi \Bigr >
\prod_{1 \le \alpha < \beta \le d} \left(
1-\frac{\eta^{2}}{(x_{i_{\alpha}}- x_{i_{\beta}})^{2} }
\right)^{-1}
  }
 \eeq
We are now in a position to use the following operator relation
(see \cite[eq. (4.2)]{TZZ15}):
\beq\label{a7a}
\det_{1\leq i,j \leq n}\left (z\delta_{ij}-\frac{\eta {\bf H}_i}{x_j-x_i+\eta}\right )=
\prod_{a=1}^{N}(z-g_a)^{{\bf M}_a}.
\eeq
The both sides are polynomials in $z$. Equating the coefficients in front of the powers of $z$, we have
 \beq\label{a7}
  \displaystyle{
\sum_{1 \le i_{1} < \ldots < i_{d} \le n}
 {\bf H}_{i_{1}}\dots {\bf H}_{i_{d}}
\prod_{1 \le \alpha < \beta \le d} \left(
1-\frac{\eta^{2}}{(x_{i_{\alpha}}- x_{i_{\beta}})^{2} } \right)^{-1}
=e_d(\{{\bf P}_j\}).
  }
 \eeq
where $e_d(\{{\bf P}_j\})$ are elementary symmetric functions defined by the generating function as
$$
\exp \left (-\sum_{k\geq 1}\frac{z^k}{k}\, {\bf P}_k\right )=
\sum_{d=0}^{n}(-1)^d z^d e_d(\{{\bf P}_j\}),
$$
and ${\bf P}_k = \sum_a {\bf M}_a g_a^k$.
In particular, for $k=1,2,3$
 \beq\label{a72}
  \displaystyle{
\begin{array}{l}
\displaystyle{
 e_1=\sum_{a=1}^{N}{\bf M}_ag_a}\,,
\\ \\
\displaystyle{
 e_2=\frac{1}{2}\Bigl (\sum_{a=1}^{N}{\bf M}_ag_a
\Bigr )^2- \frac{1}{2}\sum_{a=1}^{N}{\bf M}_ag_a^2}\,,
\\ \\
\displaystyle{
 e_3=\frac{1}{6}\Bigl (\sum_{a=1}^{N}{\bf M}_a g_a \Bigr
)^3- \frac{1}{2}\Bigl (\sum_{a=1}^{N}{\bf M}_a g_a^2 \Bigr )
 \Bigl (\sum_{b=1}^{N}{\bf M}_b g_b \Bigr ) +
\frac{1}{3} \sum_{a=1}^{N}{\bf M}_a g_a^3 }\,.
\end{array}
  }
 \eeq
 The vector $\Bigl |\Phi \Bigr >$ is an eigenvector for these operators with
 the eigenvalues given by the same formulas with ${\bf M}_a\to M_a$. Therefore,
 plugging (\ref{a7}) into (\ref{a6}) we get $\hat {\cal
H}_d\Psi=E_d\Psi$ with the eigenvalue
 \beq\label{a8}
  \displaystyle{
E_d=e_d(\{\sum_a M_a g_a^k\})\,.
  }
 \eeq
 It is easy to see that the right hand side is the elementary symmetric polynomial of $n$ variables
 $e_d(\underbrace{g_1, \ldots , g_1}_{M_1},\, \ldots \,
\underbrace{g_N, \ldots , g_N}_{M_N})$.

\section{The qKZ-Ruijsenaars correspondence in the tri\-go\-no\-met\-ric case}

The trigonometric (hyperbolic) analog of the $R$-matrix (\ref{r1}) which participates
in the qKZ equations (\ref{qkz1}) reads
 \beq\label{tr1}
{\bf R}(x)\! =\! \sum_{a=1}^N e_{aa}\otimes e_{aa}+\frac{\sinh
x}{\sinh (x\! +\! \eta )} \sum_{a\neq b}^n e_{aa}\otimes e_{bb}+
\frac{\sinh \eta}{\sinh (x\! +\! \eta )}\sum_{a<b}^n \Bigl ( e^x
e_{ab}\otimes e_{ba}+e^{-x}e_{ba}\otimes e_{ab}\Bigr ).
 \eeq
(In this section we use the same notation for analogous objects in the trigonometric
and rational cases.) After some algebra it can be represented in the form
 \beq\label{tr2}
{\bf R}_{12}(x)={\bf P}_{12}+ \frac{\sinh x}{\sinh (x+\eta )} \,
\Bigl ({\bf I}-{\bf P}^q_{12} \Bigr ),
 \eeq
where
 $$
{\bf P}^q_{12} =\sum_{a=1}^N e_{aa}\otimes e_{aa} +q\sum_{a>b}^n
e_{ab}\otimes e_{ba} +q^{-1}\sum_{a<b}^n e_{ab}\otimes e_{ba}\,,
\qquad q=e^{\eta},
 $$
is the $q$-permutation operator acting as follows:
 \beq\label{tr3}
{\bf P}_{12}^q\, e_a\otimes e_b =\left \{
\begin{array}{l}\phantom{^{-1}}q e_b\otimes e_a, \quad a<b
\\  q^{-1} e_b\otimes e_a, \quad a>b
\\  \phantom{^{-1}}\,\,\, e_b\otimes e_a, \quad a=b
\end{array}
\right.
 \eeq
Note that ${\bf P}^q_{ij}={\bf P}^{1/q}_{j\, i}$.
The unitarity condition can be easily checked.

We also introduce the $R$-matrix
 \beq\label{tr4}
{\widetilde {\bf R}}_{12}(x)= \frac{\sinh (x+\eta )}{\sinh x}\, {\bf
R}_{12}(x)= {\bf I}-{\bf P}_{12}^q + \frac{\sinh (x+\eta )}{\sinh
x}\, {\bf P}_{12}
 \eeq
and the transfer matrix $ {\bf T}(x)=\mbox{tr}_0 \Bigl ( \widetilde
{\bf R}_{0n}(x-x_n)\ldots \widetilde {\bf R}_{01}(x-x_1)\, {\bf
g}^{(0)}\Bigr ) $ of the inhomogeneous XXZ spin chain with twisted
boundary conditions. Similarly to the rational case, the commuting
Hamiltonians are defined by the pole expansion
 $$
{\bf T}(x)={\bf C}+\sinh \eta \sum_{k=1}^n {\bf H}_k \coth (x-x_k),
\qquad {\bf H}_k=(\sinh \eta )^{-1}\,\mbox{res}_{x=x_k}{\bf T}(x).
 $$
They are expressed through the $R$-matrices by the same formula (\ref{r6}).
We have (see \cite{BLZZ16}):
$\displaystyle{
{\bf T}(\pm \infty )={\bf C}\pm \sinh \eta \sum_k {\bf H}_k =\sum_{a=1}^{N}
g_ae^{\pm \eta {\bf M}_a},}
$
where ${\bf C}$ is some
operator and the weight operators ${\bf M}_a$ are defined in the same way as before. Hence
 \beq\label{tr6}
\sum_{k=1}^n {\bf H}_k = \sum_{a=1}^{N}g_a  \frac{\sinh (\eta {\bf M}_a)}{\sinh \eta}\,.
 \eeq
Similarly to (\ref{r5}), we have:
 \beq\label{tr6a}
{\bf H}_i={\bf K}_i^{(0)}\prod_{j\neq i}^n \frac{\sinh(x_i-x_j+\eta )}{\sinh (x_i-x_j)}.
 \eeq

As in the rational case, the trigonometric operators ${\bf
K}_i^{(\hbar )}$ respect the weight decomposition (\ref{weight1}).
Let $\displaystyle{\Bigl |\Phi \Bigr >=\sum_J \Phi_J \Bigl |J\Bigr
>}$ be any solution of the trigonometric qKZ equations belonging to
the weight subspace ${\cal V}(\{M_a \})$. Let $\ell (J)$ be the
minimal number of elementary permutations $\sigma_{i\,i\! +\! 1}\in
S_n$ which are required to get the multi-index $J=(j_1, j_2, \ldots
, j_n)$ from the ``minimal''
 one\footnote{Put it differently, any $\Bigl|J\Bigr>$ can
be constructed from the minimal one
 $\displaystyle{\Bigl|J_{min}\Bigr>=e_1^{\otimes M_1}\otimes...\otimes e_N^{\otimes M_N}}$ in $\ell(J)$ steps by
 applying elementary permutations of neighboring tensor components $\sigma_{k_i\,k_i\! +\!
1}$:
  $$
 \displaystyle{
 \Bigl|J_{min}\Bigr>=\Bigl|J^{(0)}\Bigr>\stackrel{\sigma_{k_1\,k_1\! +\!
1}}{\longrightarrow} \Bigl|J^{(1)}\Bigr>\stackrel{\sigma_{k_2\,k_2\!
+\! 1}}{\longrightarrow}
\Bigl|J^{(2)}\Bigr>\stackrel{\sigma_{k_3\,k_3\! +\!
1}}{\longrightarrow}...\stackrel{\sigma_{k_{\ell (J)}\,k_{\ell
(J)}\! +\! 1}}{\longrightarrow} \Bigl|J^{({\ell
(J)})}\Bigr>=\Bigl|J\Bigr>\,,
 }
 $$
where $ \displaystyle{
\Bigl|J^{(i)}\Bigr>=e_{j_1^{(i)}}\otimes...\otimes e_{j_n^{(i)}} }$
and each time $j_{k_i}^{(i-1)}<j_{k_i+1}^{(i-1)}$.},
 where the $j_k$'s are ordered as $1\leq j_1\leq j_2\leq
\ldots \leq j_n\leq N$.
 In the case $n=N$, $M_1=M_2=\ldots =M_N=1$
the $\ell (J)$ is what is called length of the permutation
$(12\ldots N)\to (j_1j_2\ldots j_N)$. We claim that the function
 \beq\label{tr7}
\Psi =\sum_J q^{\ell (J)}\Phi_J
 \eeq
is an eigenfunction of the trigonometric
Macdonald operator $\hat {\cal H}$ with the eigenvalue
$\displaystyle{E=\sum_{a=1}^N g_a}\frac{\sinh (\eta M_a)}{\sinh \eta}\,$:
 \beq\label{tr8}
\sum_{i=1}^n  \prod_{j\neq i}^n
\frac{\sinh (x_i\! -\! x_j \! +\! \eta )}{\sinh (x_i -x_j)}\, \Psi (x_1, \ldots , x_i\! +\! \eta \hbar ,
\ldots , x_n) =E\Psi (x_1, \ldots , x_n).
 \eeq

The idea of the proof is the same as in the rational case.
We consider the covector
 $$
\Bigl < \Omega _q\Bigr |=\sum_J q^{\ell (J)}\Bigl < J \Bigr |\, ,
 $$
then $\Psi =\Bigl <\Omega _q\Bigr | \Phi \Bigr >$. It is not
difficult to see that $\Bigl <\Omega_q \Bigr |{\bf P}^q_{i,i\! -\!
1}=\Bigl <\Omega_q \Bigr |$.\footnote{Indeed, from (\ref{tr3}) and
construction of $\Bigl|J\Bigr>$ from $\Bigl|J_{min}\Bigr>$ one gets
${\bf P}^q_{i\! -\! 1,i}\Bigl|\Omega_q \Bigr
>=\Bigl|\Omega_q \Bigr >$. The matrix transposition of this equality gives
$\Bigl <\Omega_q \Bigr |=\Bigl <\Omega_q \Bigr |\Bigl({\bf P}^q_{i\!
-\! 1,i}\Bigr)^{\!T}=\Bigl <\Omega_q \Bigr |\,{\bf P}^{1/q}_{i\! -\!
1,i}=\Bigl <\Omega_q \Bigr |\,{\bf P}^{q}_{i,i\! -\! 1}$.} This
implies the important relation
 \beq\label{tr9}
\Bigl <\Omega_q \Bigr |\, {\bf R}_{i\, i\! -\! 1}(x)=\Bigl <\Omega_q \Bigr |\, {\bf P}_{i\, i\! -\! 1},
\qquad i=2, \ldots , n
 \eeq
(the second term in (\ref{tr2}) disappears). Using the relation ${\bf P}_{i\, i\! -\! 1}
{\bf P}^q_{i\, i\! -\! 2}={\bf P}^q_{i\! -\! 1 \, i\! -\! 2}{\bf P}_{i\, i\! -\! 1}$, one can show
by induction that
 $$
\Bigl <\Omega_q\Bigr |\, {\bf R}_{i\, i\! -\! 1}(x_i \! -\! x_{i-1}\! +\! \eta \hbar )
\ldots {\bf R}_{i1}(x_i \! -\! x_{1}\! +\! \eta \hbar )=
\Bigl <\Omega_q\Bigr |\, {\bf P}_{i\, i\! -\! 1}\ldots {\bf P}_{i1}
 $$
for all $i=2, \ldots , n$. Since the right hand side does not depend on the spectral parameters,
we can substitute each $R$-matrix in the left hand side by the same one with $\hbar =0$ and
conclude that $\Bigl <\Omega _q\Bigr |{\bf K}_{i}^{(\hbar )}=
\Bigl <\Omega_q \Bigr |{\bf K}_{i}^{(0 )}$. The projection of the $i$-th qKZ equation
onto the covector $\Bigl <\Omega _q\Bigr |$ reads
 $$
e^{\eta \hbar \p_{x_i}}\Bigl <\Omega _q\Bigr |\Phi \Bigr >=
e^{\eta \hbar \p_{x_i}}\Psi =
\Bigl <\Omega_q \Bigr |   {\bf K}_{i}^{(\hbar )}\Bigl | \Phi \Bigr >=
\Bigl <\Omega_q \Bigr |   {\bf K}_{i}^{(0 )}\Bigl | \Phi \Bigr >.
 $$
In the same way as in the rational case, we
multiply this by $\displaystyle{\prod_{j\neq i}^n \frac{\sinh (x_i-x_j +\eta )}{\sinh (x_i-x_j)}}$
and sum over $i$ to get:
 $$
\sum_{i=1}^n \left ( \prod_{j\neq i}^n \frac{\sinh (x_i-x_j +\eta )}{\sinh (x_i-x_j)}\right )
e^{\eta \hbar \p_{x_i}}\Psi \,\, =\,\,
\sum_{i=1}^n  \prod_{j\neq i}^n \frac{\sinh (x_i-x_j +\eta )}{\sinh (x_i-x_j)}\,
\Bigl <\Omega_q \Bigr |   {\bf K}_{i}^{(0 )}\Bigl | \Phi \Bigr >
 $$
 $$
=\, \sum_{i=1}^n \Bigl <\Omega_q \Bigr |   {\bf H}_{i}\Bigl | \Phi \Bigr > \,\, =
\,\,
\sum_{a=1}^N g_a \Bigl <\Omega _q\Bigr | \frac{\sinh (\eta  {\bf M}_a)}{\sinh \eta}
\Bigl | \Phi \Bigr >\,\, =\,\,
\left (\sum_{a=1}^N g_a \frac{\sinh (\eta M_a )}{\sinh \eta}\right )\Psi .
 $$
Here we have used (\ref{tr6}) and (\ref{tr6a}).

Similarly to the rational case, the correspondence can be extended to the higher
trigonometric Macdonald-Ruijsenaars operators.

\section{Conclusion and discussion}

We have established the correspondence between solutions to the qKZ equations (\ref{qkz1})
in different weight subspaces of $V^{\otimes n}$
and solutions to the spectral problem for the $n$-body Ruijsenaars model with Planck's
constant $\hbar$, with
$V$ being the space of $N$-dimensional vector representation of $GL(N)$.
The proof appears to be even simpler than in the case of
the correspondence between the differential KZ equations and the quantum Calogero model
(see \cite{Matsuo,Cherednik,FV} and \cite{ZZ}).
In the limit $\hbar \to 0$ we obtain
the quantum-classsical correspondence
\cite{AKLTZ13,GZZ14,TZZ15} between
the quantum spin chain (XXX or XXZ) and the classical Ruijsenaars system of particles \cite{RS}
(rational or trigonometric).

The Hamiltonian of the classical Ruijsenaars system has the form
 \beq\label{c1}
{\cal H}=\sum_{i=1}^n e^{\eta p_i}
\prod_{j\neq i}^n \frac{x_i-x_j +\eta}{x_i-x_j}
 \eeq
with the usual Poisson brackets $\{p_i, x_j\}=\delta_{ij}$ (for simplicity we consider
the rational case).
The model is integrable, with the Lax matrix
 $$
L_{ij}=\frac{\dot x_j}{x_i-x_j+\eta},
 $$
where
 \beq\label{dotx}
\dot x_i= \frac{\p {\cal H}}{\p p_i}=\eta e^{\eta p_i}
\prod_{j\neq i}^n \frac{x_i-x_j +\eta}{x_i-x_j}
 \eeq
is the velocity of the $i$-th particle.
The higher Hamiltonians in involution are coefficients of the characteristic polynomial of the
Lax matrix:
$$
\det_{ij} (z\delta_{ij}-L_{ij})= \sum_{d=0}^{n}
(-1)^d z^{n-d}{\cal H}_d, \qquad {\cal H}_1={\cal H}
$$

The correspondence with the quantum spin chain goes as follows.
Let the eigenvalues of the $n\! \times \! n$ Lax matrix be the twist parameters
$g_a$ with multiplicities $M_a$ (recall that $M_1 +\ldots +M_a=n$). This means that we
consider the level set of all the classical Hamiltonians
 $$
{\cal H}_d =
e_d(\underbrace{g_1, \ldots , g_1}_{M_1},\, \ldots \,
\underbrace{g_N, \ldots , g_N}_{M_N})
\quad d\geq 1, \quad M_a\in \ZZ_{\geq 0},
 $$
with fixed coordinates $x_i$.
Then the admissible values of velocities, $\dot x_i$, are equal to
$\eta H_i$, where the $H_i$'s are
eigenvalues of the spin chain Hamiltonians ${\bf H}_i$ in the weight subspace
${\cal V}(\{M_a\})$ for the model
with the inhomogeneity parameters $x_i$ and the twist matrix
${\bf g}=\mbox{diag}\, (g_1, \ldots , g_N)$. In fact the admissible values of $\dot x_i$'s obey
a system of algebraic equations (see \cite{TZZ15}). Different solutions of this system correspond to
different eigenstates of the spin chain Hamiltonians.

In the trigonometric case eigenvalues of the Lax matrix
$\displaystyle{
L^{\rm trig}_{ij}=\frac{\dot x_j}{\sinh (x_i-x_j +\eta )}}
$
should form ``multiplicative strings'' of lengths $M_a$ centered at $g_a$:
 $$
g_a^{(\alpha )}=g_ae^{-(M_a-1)\eta +2\eta \alpha}, \qquad \alpha = 0,1, \ldots , M_a-1.
 $$
Then $\dot x_i/\eta$ are eigenvalues of the XXZ spin chain Hamiltonians.
The eigenvalue $E=\displaystyle{
\sum_{a=1}^N g_a \frac{\sinh (\eta M_a )}{\sinh \eta}}$ of the Ruijsenaars operator
agrees with this since it is clear that
 $$
E=\sum_{a=1}^n \sum_{\alpha =0}^{M_a-1}g_a^{(\alpha )}.
 $$

Comparing this with (\ref{quasi1}) and taking into account (\ref{r5}), (\ref{dotx}), we see that
in the limit $\hbar \to 0$ of the qKZ system (which is the spectral problem for
${\bf H}_i$'s or ${\bf K}_i^{(0)}$'s) the other side of the correspondence becomes the
classical Ruijsenaars model. In other words, we can say that
the quantization of the classical Ruijsenaars system of particles
with the Planck constant $\hbar$
($p_i \to \hbar \p_{x_i}$) corresponds to the passage from the spectral problem
for the spin chain to the system of qKZ equations with the step parameter $\eta \hbar$.

\section*{Acknowledgments}

We thank A.Liashyk for discussions. The work of A. Zabrodin has been
funded by the Russian Academic Excellence Project `5-100'. It was
also supported in part by RFBR grant 15-01-05990 and by joint RFBR
grant 15-52-50041 YaF$_a$. The work of A. Zotov was supported by
RFBR grant 15-01-04217 and by joint RFBR project 15-51-52031
HHC$_a$.

\end{document}